\documentclass[preprint,aps,prx,amsmath,amssymb]{revtex4-1}
\usepackage{bm}
\usepackage{graphicx}
\usepackage{times}
\usepackage{newtxtext}

\newcommand{\wt}[1]{\widetilde{#1}}

\newcommand{\mean}[1]{\left< {#1} \right>} % Displaystyle
\newcommand{\tmean}[1]{\langle {#1} \rangle} % For non-displaystyle (text) angles

\newcommand{\Figref}[1]{Fig.~\ref{#1}}
\newcommand{\Figsref}[1]{Figs.~\ref{#1}}
\newcommand{\Secref}[1]{Sec.~\ref{#1}}
\newcommand{\Tabref}[1]{Table~\ref{#1}}

\newcommand{\Sn}{\Phi} % Shot noise process
\newcommand{\Snw}{\varphi} % Shot noise waveform

\newcommand{\Snn}{\Psi}
\newcommand{\nSnn}{\wt{\Psi}}
\newcommand{\mSnn}{\mean{\Psi}}
\newcommand{\rSnn}{{\Psi_\rms}}

\newcommand{\mA}{\left< A \right>} %Mean of A
\newcommand{\mSn}{\left< \Sn \right>} % Mean of signal
\newcommand{\rSn}{\Sn_{\text{rms}}} % RMS of signal
\newcommand{\rms}{\text{rms}}
\newcommand{\td}{\tau_{\text{d}}} %duration time
\newcommand{\tw}{\tau_{\text{w}}} % Average waiting time
\begin{document}

\title{Universality of Poisson-driven plasma fluctuations in the Alcator C-Mod scrape-off layer}

\author{A.~Theodorsen}
\email{audun.theodorsen@uit.no}
\author{O.~E.~Garcia}
\email{odd.erik.garcia@uit.no}
\author{R.~Kube}
\affiliation{Department of Physics and Technology, UiT The Arctic University of Norway, N-9037 Troms{\o}, Norway}
\author{B.~La{B}ombard}
\author{J.~L.~Terry}
\affiliation{MIT Plasma Science and Fusion Center, Cambridge, MA 02139, United States of America}

\date{\today}

\begin{abstract}
%Gas puff imaging data time series from the scrape-off layer in the Alcator C-Mod device have been analyzed for various plasma parameters and confinement modes. The large-amplitude fluctuations can be described as a super-position of pulses with fixed shape and constant duration. By applying a deconvolution algorithm on the data time series with a two-sided exponential pulse function, the arrival times and amplitudes of the pulses can be estimated and the measurement time series can be reconstructed with high accuracy. The pulse amplitudes are shown to follow an exponential distribution. The waiting times between pulses are uncorrelated, their distribution has an exponential tail, and the number of arrivals is a linear function of time. This shows that pulse arrivals follow a homogeneous Poisson process. Identical statistical properties applies to both ohmic and high confinement mode plasmas, clearly demonstrating universality of the fluctuations in the boundary region of Alcator C-Mod.

Large-amplitude, intermittent fluctuations are ubiquitous in the boundary region of magnetically confined plasmas and lead to detrimental plasma-wall interactions in the next-generation, high duty cycle fusion power experiments. Using gas puff imaging data time series from the scrape-off layer in the Alcator C-Mod device, it is here demonstrated that the large-amplitude fluctuations can be described as a super-position of pulses with fixed shape and constant duration. By applying a new deconvolution algorithm on the data time series with a two-sided exponential pulse function, the arrival times and amplitudes of the pulses can be estimated and the measurement time series can be reconstructed with high accuracy. The pulse amplitudes are shown to follow an exponential distribution. The waiting times between pulses are uncorrelated, their distribution has an exponential tail, and the number of arrivals is a linear function of time. This demonstrates that pulse arrivals follow a homogeneous Poisson process. Identical statistical properties apply to both ohmic and high confinement mode plasmas, clearly demonstrating universality of the fluctuation statistics in the boundary region of Alcator C-Mod.
\end{abstract}

\maketitle

\section{Introduction}
Extensive scientific investigations have revealed that cross-field transport of particles and heat in the scrape-off layer (SOL) of magnetically confined plasmas is caused by radial motion of blob-like filament structures \cite{dippolito-2004,zweben-2007,krasheninnikov-2008,garcia-2009,dippolito-2011}. This poses several challenges for future magnetic fusion energy reactors, including enhanced erosion rates of the main chamber walls \cite{labombard-2001,pitts-2005,lipschultz-2007,dippolito-2008,marandet-2016}. There is also strong evidence that the turbulence-driven cross-field transport is related to the empirical discharge density limit \cite{labombard-2005,antar-2005,dippolito-2006,garcia-ppcf-2007,guzdar-2007}. The fluctuation-induced transport and associated plasma--wall interactions evidently depend on the amplitude of the filaments and their frequency of occurrence \cite{garcia-prl-2012,garcia-pop-2016,militello-nf-2016}.

Radial motion of blob-like structures results in single-point recordings dominated by large-amplitude bursts. Recently, a stochastic model was introduced, describing the fluctuations as a super-position of uncorrelated pulses with an exponential shape and constant duration \cite{garcia-prl-2012,garcia-pop-2016,militello-nf-2016,militello-ppcf-2016,theodorsen-ppcf-2018}. Predictions of this model, including the probability density function and the frequency power spectral density, are in excellent agreement with Langmuir probe and gas puff imaging (GPI) measurements obtained in ohmic and low confinement modes (L-modes) of several tokamak devices \cite{graves-2005,kube-2016,theodorsen-ppcf-2016,garcia-nme-2017,theodorsen-nf-2017,garcia-pop-2018,kube-ppcf-2018,walkden-ppcf-2017}.

In this paper, a new method is introduced in order to reveal the pulse amplitudes and arrival times directly, without inferring their properties from the predictions of the model. This is achieved by reformulating the stochastic model as a convolution of the pulse function with a train of delta pulses and invoking a deconvolution algorithm. Applying this method to  measurement data from GPI of the SOL in the Alcator C-Mod device, it is for the first time demonstrated that the pulses occur according to a Poisson process and that the pulse amplitudes are exponentially distributed. These statistical properties are identical for both ohmic and high confinement modes (H-modes), providing further evidence for universality of the statistical properties of the fluctuations in the boundary region of magnetically confined plasmas. The results presented here complement and extend previous work that pointed out similarities between SOL plasma fluctuations in L- and H-modes \cite{rudakov-2002,antar-2008,ionita-2013,zweben-2015,zweben-2016,garcia-pop-2018}. In particular, it extends the work in \cite{garcia-pop-2018} by using the new deconvolution method, and the results presented in this contribution should be compared to the conditional averaging performed in \cite{garcia-pop-2018}. 

\section{Experimental setup}
All experiments analyzed here were deuterium fuelled plasmas in a lower single null divertor configuration. The GPI diagnostic on Alcator C-Mod consists of a $9\times10$ array of toroidal views of a localized gas puff \cite{cziegler-2010}. The spot size of the horizontal lines-of-sight are $3.8\,\text{mm}$ in diameter at the gas cloud. The views are brought via optical fibers to high sensitivity avalanche photo diodes and the signals are digitized at a rate of $2\times10^6$ frames per second. In this study, the He I line emission from the localized He gas puff is investigated for a view position in the far SOL with major radius $R=90.69\,\text{cm}$ and vertical position $Z=-2.99\,\text{cm}$, which is $1.0$ to $1.8\,\text{cm}$ outside the last closed magnetic flux surface for the cases studied here.
%, that is locally enhanced in the object plane by an extended He gas puff from a nearby nozzle, is investigated for a diode view position in the far SOL with major radius $R=90.69\,\text{cm}$ and vertical position $Z=-2.99\,\text{cm}$.

We will investigate time series from the GPI diagnostic for various plasma parameters and confinement modes as listed in \Tabref{tab:data-details}. All time series have a duration of $100\,\text{ms}$, and these intervals have been chosen such that the time series are approximately stationary without using moving averages or filtering. Two ohmically heated plasma states are analyzed, one low density case denoted `lO' with a Greenwald density fraction of $\overline{n}_\text{e}/n_\text{G}=0.3$ and one high density case denoted `hO' with a Greenwald fraction of $0.6$. Here $\overline{n}_\text{e}$ is the line-averaged electron density and the Greenwald density is given by $n_\text{G}=(I_\text{p}/\pi a^2)10^{20}\,\text{m}^{-3}$, where the plasma current $I_\text{p}$ is given in units of MA and the minor radius $a$ is in units of meters \cite{greenwald-1988}.

\begin{table}
    \centering
    \begin{tabular}{ c | c | c | c | c | c | c}
        Plasma state & Shot number & $t_0/\text{s}$ & $n_\text{e}/n_\text{G}$ & $B_0 /\text{T}$ & $I_\text{p} / \text{MA}$ & $P_\text{RF} / \text{MW}$ \\
        \hline
        lO & 1150618021 & 0.80 & 0.3 & 4.1 & 0.6 & 0 \\
        hO & 1150618036 & 0.74 & 0.6 & 4.1 & 0.6 & 0 \\
        qH & 1110201011 & 1.13 & 0.5 & 5.4 & 1.2 & 3.0 \\
        eH & 1110201016 & 1.23 & 0.6  & 5.4 & 0.9 & 3.0 \\
    \end{tabular}\\
    \caption{\label{tab:data-details} Notation and shot number for the discharges analyzed here. `lO' denotes a low density Ohmic plasma, `hO' a high density Ohmic state, `qH' a quiescent H-mode and `eH' an enhanced D-alpha H-mode. Each time series analyzed has a duration of $100\,\text{ms}$ and $t_0$ gives the starting time. The other columns give the Greenwald fraction of the line-averaged density, the magnetic field on axis, the plasma current, and the ICRF heating power.}
\end{table}

\begin{figure}
  \centering
    \includegraphics[width=8.5cm]{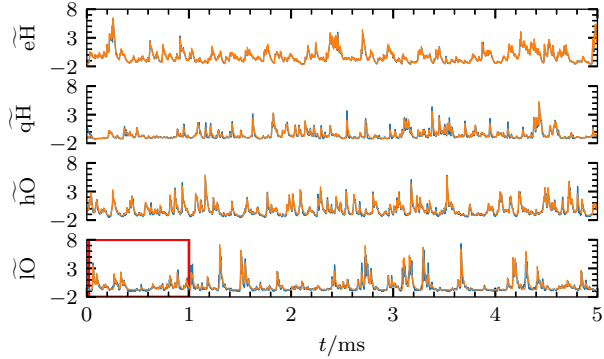}
    \caption{\label{fig:data-fit}Excerpt of GPI measurement data time series for four different plasma states (blue lines). Also shown are reconstructed time series from the deconvolution algorithm (orange lines). The red box indicates the exerpt presented in \Figref{fig:detail-data-fit}.}
\end{figure}

In the case of strong ion cyclotron range of frequencies (ICRF) heating, there are two different types of H-modes on Alcator C-Mod without edge localized modes (ELMs). One is the enhanced D-alpha H-mode, here denoted `eH', which is a steady mode of operation with an edge transport barrier. A quasi-coherent mode in the edge region prevents impurities from accumulating in the core, resulting in a steady state H-mode without ELMs \cite{labombard-2014}. Another type of ELM-free H-mode on Alcator C-Mod is the so-called quiescent H-mode. In this case there is a strong particle and heat transport barrier but a lack of macroscopic instabilities of the edge pedestal. This results in an accumulation of impurities in the core, which eventually causes a radiative collapse of the plasma. Such a state, here denoted `qH', with stationary far SOL plasma parameters has also been analyzed here. A short part of these GPI data time series are presented in \Figref{fig:data-fit}, demonstrating the intermittent nature of the fluctuations for all plasma parameters and confinement modes. Here, the time series are normalized by subtracting the mean value and dividing by the rms-value, $\widetilde{D} = (D-\mean{D})/D_\rms$, where $D$ denotes any of the GPI data time series.

\section{Fluctuation statistics}
In previous work, the predictions of the filtered Poisson process (FPP) have been shown to be in excellent agreement with analysis of experimental measurement data from the SOL of numerous tokamak experiments. The FPP is given by a super-position of uncorrelated pulses \cite{garcia-prl-2012,kube-2015,theodorsen-pop-2016,garcia-pop-2016,militello-nf-2016,militello-ppcf-2016,theodorsen-ps-2017,garcia-pop-2017-2,theodorsen-pre-2018,theodorsen-ppcf-2018},
\begin{equation}\label{eq:fpp}
    \Sn_K(t) = \sum_{k=1}^{K(T)} A_k \Snw\left( \frac{t - t_k}{\td} \right)
\end{equation}
on the interval $t\in[0,T]$, where $T$ is the full time duration of the signal. All pulses have the same pulse duration time $\td$. The pulse arrival times $t_k$ are independently and uniformly distributed on $[0,T]$. Correspondingly, $K(T)$ is a Poisson process with intensity $T/\tw$ and the waiting times are exponentially distributed with mean value $\tw$. The amplitudes $A_k$ are taken to be independent and exponentially distributed with mean value $\mA$. The pulse function $\Snw$ is given by a two-sided exponential function
\begin{equation}\label{eq:pulse-shape}
    \Snw(x) = \begin{cases}
        \exp\left( -x/(1-\lambda) \right), \quad  &x \geq 0, \\
        \exp\left( x/\lambda \right), &x < 0,
    \end{cases}
\end{equation}
where $x$ is a unitless variable and $\lambda$ is the pulse asymmetry parameter restricted to the range $\lambda \in [0,1]$. %
The most important parameter describing this process is the intermittency parameter $\gamma=\td/\tw$, which determines the degree of pulse overlap \cite{garcia-prl-2012}.

It can be shown that the stationary probability density function (PDF) of the FPP with two-sided exponential pulses is a Gamma distribution with the shape parameter $\gamma$ and scale parameter $\mA$ \cite{garcia-pop-2016},
\begin{equation}\label{eq:pdf-fpp}
    P_\Sn(\phi) = \frac{\phi^{\gamma-1}}{\mA^\gamma \Gamma(\gamma)} \exp\left( -\frac{\phi}{\mA} \right),\, \phi > 0.
\end{equation}
The four lowest order moments of $\Sn$ are given by the mean $\mSn=\gamma\mA$, the variance $\rSn^2=\gamma\mA^2$, the skewness $S_\Sn=2/\gamma^{1/2}$ and the flatness $F_\Sn=3+6/\gamma$.

In order to account for measurement noise and small discrepancies from the pure two-sided exponential pulse function, we introduce a normally distributed noise signal $X(t)$, with mean value $\mu$, variance $X_\rms^2 = \epsilon \Sn_\rms^2$ and the same power spectral density as $\Sn(t)$ \cite{theodorsen-ps-2017,theodorsen-ppcf-2018}. The \emph{noise parameter} $\epsilon$ is defined as
\begin{equation}
    \epsilon = \frac{X_\rms^2}{\Sn_\rms^2}.
\end{equation}
We denote the sum of the FPP with noise as 
\begin{equation}
\Snn(t) = \Sn(t) + X(t).
\end{equation} 
The distribution of $\Snn$ is a convolution between a Gamma distribution and a normal distribution, and the first four moments are given by $\mSnn = \mu + \gamma \mA$, $\rSnn^2 = (1+\epsilon)\gamma \mA^2$, $S_\Snn = 2 \left[ (1+\epsilon)^3 \gamma  \right]^{-1/2}$ and $F_\Snn = 3+ 6 (1+\epsilon)^{-3} \gamma^{-1}$ \cite{theodorsen-ps-2017}.

\begin{figure}
  \centering
    \includegraphics[width=8.5cm]{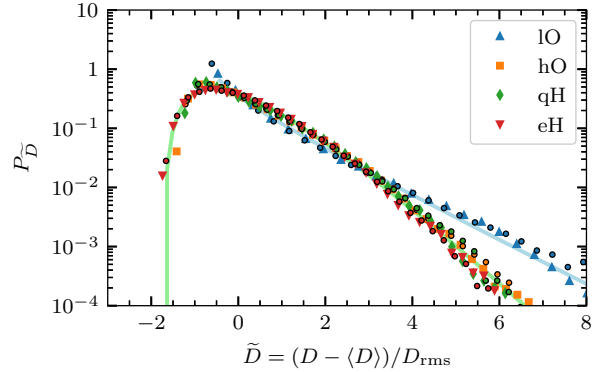}
    \caption{\label{fig:pdf}PDFs of GPI measurement data (symbols), the corresponding distributions from the reconstructed time series (same color but outlined circles) described in \Secref{sec:deconv-result}, and two Gamma distributions with shape parameters $\gamma=2/3$ (light blue line) and $\gamma=3$ (light green line). $D$ denotes any of the original or reconstructed time series.}
\end{figure}

Normalizing $\Snn$ by subtracting the mean and dividing by the rms-value, $\nSnn = (\Snn-\mSnn)/\rSnn$, eliminates $\mu$ and $\mA$ as explicit parameters. In \Figref{fig:pdf}, the PDFs of the measurement data are compared to the Gamma distribution with shape parameters $\gamma=2/3$ and $\gamma=3$. 
By using the method described in \cite{theodorsen-ppcf-2018}, $\gamma$ and $\epsilon$ can be estimated from the empirical characteristic function of the normalized GPI time series. Using these values and the first two moments of the time series, $\mu$ can be estimated as $\mu = \mSnn - \sqrt{\gamma/(1+\epsilon)} \rSnn$. 

The estimated parameters are presented in \Tabref{tab:est-par} along with the mean value of the time series. Consistent with \Figref{fig:data-fit}, the low density Ohmic state is strongly intermittent, while pulse overlap is more significant for the enhanced D-alpha H-mode state as expected from the moment estimation. In all cases, $\epsilon$ is very moderate, or practically vanishing, consistent with the good agreement between the data and a pure Gamma distribution in \Figref{fig:pdf}. In all cases $\mu/\mSnn$ ranges from 0.2-0.4, indicating that the mean value consists primarily, but not exclusively, of the mean value of the pulses.

\begin{table}
    \centering
    \begin{tabular}{ c | c | c | c | c}
        Plasma state & $\gamma$ & $\epsilon$ & $\mu$ & $\mSnn$ \\
        \hline
        lO & 0.60 & 0.02 & 0.06 & 0.14 \\
        hO & 1.71 & 0.01 & 0.08 & 0.19 \\
        qH & 1.51 & 0.00 & 0.11 & 0.45 \\
        eH & 3.30 & 0.00 & 0.24 & 0.84 \\
    \end{tabular}\\
    \caption{\label{tab:est-par} Estimated parameters of the GPI time series.}
\end{table}

The pulse parameters $\td$ and $\lambda$ can be estimated from the power spectral density and the conditionally averaged pulse shape of a time series. In \cite{garcia-pop-2018}, it was found that a pulse shape with $\td = 20\, \mu s$ and $\lambda = 1/10$ describes the power spectral density and conditional average of all data time series presented here well. These results are presented here to complete the parameter estimation.

\begin{figure}
  \centering
    \includegraphics[width=8.5cm]{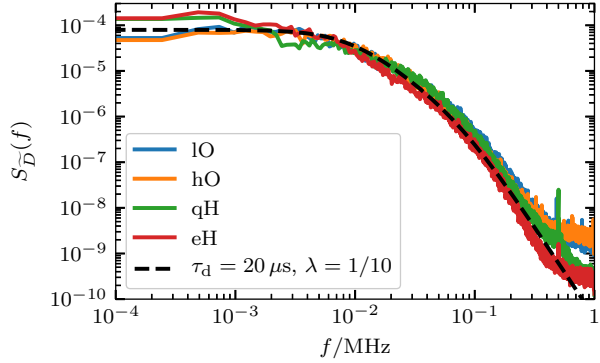}
    \caption{\label{fig:psd} Power spectral density of GPI measurement data. The black dashed line gives the prediction from the stochastic model.}
    % The broken line shows the conditional average of synthetic data from a realization of the FPP.}
\end{figure}

In \Figref{fig:psd}, the power spectral densities for the four different plasma states are presented together with the analytic prediction of the power spectrum of an FPP with additive noise \cite{theodorsen-ps-2017}. The power spectra show a remarkable similarity and agrees with the prediction from the stochastic model using $\td = 20\, \mu \text{s}$ and $\lambda = 1/10$. The universality of the power spectra from GPI time series from the SOL of Alcator C-Mod for different line-averaged densities, confinement regimes and at different radial positions in the SOL has been noted before \cite{theodorsen-nf-2017,garcia-pop-2018}.

\begin{figure}
  \centering
    \includegraphics[width=8.5cm]{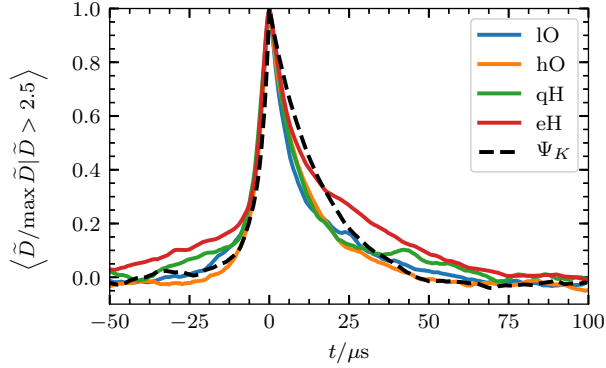}
    \caption{\label{fig:cond_av}Conditionally averaged waveform of GPI measurements and synthetic data for fluctuation amplitudes larger than 2.5 times the root mean square value.}
    % The broken line shows the conditional average of synthetic data from a realization of the FPP.}
\end{figure}

In order to verify the deconvolution method, we will employ it on a synthetically generated FPP with additive noise with parameters $\gamma=1.5$, $\td = 20\,\mu \text{s}$, $\lambda=1/10$ and $\epsilon = 0.01$ in addition to the GPI data. In the following, this realization will be denoted $\Snn_K$. In \Figref{fig:cond_av}, the conditionally averaged waveform for the four different plasma states are presented together with the conditional average of $\Snn_K$. The conditional average of the synthetic signal conforms well to the general shape of the conditional average of the data time series. The somewhat longer duration time of the enhanced D-alpha H-mode case was discussed in \cite{garcia-pop-2018}, and is not taken to be significant for the purposes of the deconvolution. 
Indeed, as the RL-deconvolution is robust to small deviations in the pulse shape, we will use $\td=20\,\mu\text{s}$ and $\lambda=1/10$ as the pulse parameters for the deconvolution of all time series. Different pulse parameters have been tested, without significant deviations in the results presented in \Secref{sec:deconv-result}.
%ohmically heated plasma states and the quiescent H-mode. The slight discrepancy for large positive time lags for the enhanced D-alpha H-mode case may be due to more significant pulse overlap for this case and second-order effects not captured by the stochastic model. However, the RL-deconvolution is robust to small deviations in the pulse shape. We will therefore use $\td=15\,\mu\text{s}$ and $\lambda=1/10$ as the pulse parameters for both $\Snn_K$ and the deconvolution algorithm in the following.

\section{Deconvolution algorithm}\label{sec:deconv}
The FPP can be written as a convolution between the pulse function and a train of delta-function pulses \cite{theodorsen-ps-2017},
\begin{equation}\label{eq:fpp-convolve}
    \Sn_K(t) = \left[\Snw * f_K \right]\left(\frac{t}{\td}\right),
\end{equation}
where
\begin{equation}\label{eq:fK}
    f_K(t) = \sum\limits_{k=1}^{K(T)}A_k \delta\left( \frac{t - t_k}{\td} \right).
\end{equation}
The goal of this contribution is to obtain and investigate the properties of the pulse amplitudes $\{A_k\}_{k=1}^K$ and arrival times $\{t_k\}_{k=1}^K$ directly. In order to do this, we will use the Richardson-Lucy deconvolution algorithm \cite{richardson-1972,lucy-1974} with normally distributed noise \cite{witherspoon-1986,pruksch-1998,dellacqua-2007,tai-2011} to estimate $f_K(t)$. This algorithm is iterative, with the $n$'th iteration given by
\begin{equation}\label{eq:rl-deconv}
    f_K^{(n+1)} = f_K^{(n)} \frac{(D-\mu)*\widehat{\Snw}}{f_K^{(n)}*\Snw*\widehat{\Snw}},
\end{equation}
where $\widehat{\Snw}(t) = \Snw(-t)$. Here and in the following, $D$ denotes any of the GPI measurement data time series discussed above as well as the realization of $\Snn$ discussed below. The estimate of $\mu$ is presented in \Tabref{tab:est-par}. We note that this expression is independent of $X_\rms$. 
The initial guess $f_K^{(1)}$ is unimportant, and can be set as a positive constant or the measurement signal itself.

If $D-\mu$ and $f_K^{(1)}$ are positive definite, each iteration $f_K^{(n)}$ is as well. While $D$ is positive definite, and $f_K^{(1)}$ can be chosen positive definite, $D-\mu$ is not guaranteed to be positive definite. In practice, however, the noise level is small enough that using the absolute value of $D-\mu$ has no appreciable effect on the result of the deconvolution (the power contained in the negative part of $D-\mu$ is less than $1\,\%$ of the total signal power).
The algorithm converges to the least-squares solution \cite{dellacqua-2007}. The result of the iteration is a super-position of sharp Gaussian-like pulses, as the iteration gradually smooths the signal.
The arrival times are determined from the maxima of $f_K^{(n)}$. The amplitudes associated with each arrival is the integral of $f_K^{(n)}$ from the minima between the previous and current arrivals to the minima between the current and next arrival. The fit to the measurement data time series, $D_\text{fit}$, is then computed from these arrival times and amplitudes.

The maxima of $f_K^{(n)}$ are determined as the zeros in the derivative of $f_K^{(n)}$, where the derivative is computed by fitting $f_K^{(n)}$ to a second-order polynomial in a prescribed window.
The number of arrivals strongly depends on the window size. While the expected total number of events is $\mean{K}=\gamma T/\td$, for a discrete time series the expected number of time grid points containing events is $\mean{F}=N[(1-\exp(-\gamma \triangle_t/\td)]$ where $N=T/\td$ is the number of time grid points and $\triangle_t$ is the time step \cite{theodorsen-deconv}. Since the deconvolution procedure only discovers the presence of events at a given grid point, $\mean{F}$ is the correct number of events to use. We choose the window size minimizing the difference between the number of deconvolved events and $\mean{F}$. In the case of the GPI time series, the window sizes are $28.5\, \mu \text{s}$ (lO), $7.5\, \mu \text{s}$ (hO), $6.5\, \mu \text{s}$ (qH) and $5.5\, \mu \text{s}$ (eH), giving 2990, 8012, 7271 and 15507 events respectively. By comparison, conditional averaging of these time series gives several hundred events \cite{garcia-pop-2018}. For the synthetic time series, a window of $8.5\,\mu \text{s}$ was used, giving 7485 events in comparison to 196 events from conditional averaging. Increasing the window size eliminates small and sharp peaks in $f_K^{(n)}$ and consolidates close peaks. This comprises the noise handling inherent in the method.

\section{Result of deconvolution}\label{sec:deconv-result}
\begin{figure}
  \centering
    \includegraphics[width=8.5cm]{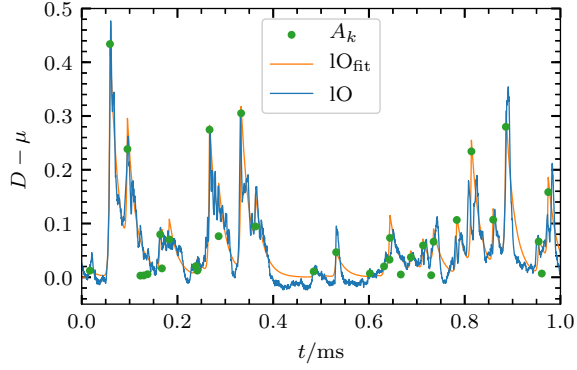}
    \caption{\label{fig:detail-data-fit}Excerpt of measurement data (blue line) and reconstructed time series (orange line) for the low density Ohmic state. The green dots show the estimated pulse arrival times and amplitudes.}
\end{figure}

The result of the deconvolution algorithm is presented in \Figref{fig:data-fit}, with the reconstructed time series plotted on top of the measurement data. In all cases, the reconstructed signal is very close to the original signal. The PDFs also correspond closely, as seen in \Figref{fig:pdf}. An excerpt of the low density Ohmic case with the reconstructed signal and pulse amplitudes is presented in \Figref{fig:detail-data-fit}. While there is some scatter around the measured signal, the reconstruction captures the main fluctuations in the GPI signal.
% It should be noted that the apparent over-estimation of some of the pulse amplitudes is not necessarily a failure of the deconvolution method since the GPI diagnostic suffers from a burn out effect for high intensity signals \cite{zweben-2017}.

\begin{figure}
  \centering
    \includegraphics[width=8.5cm]{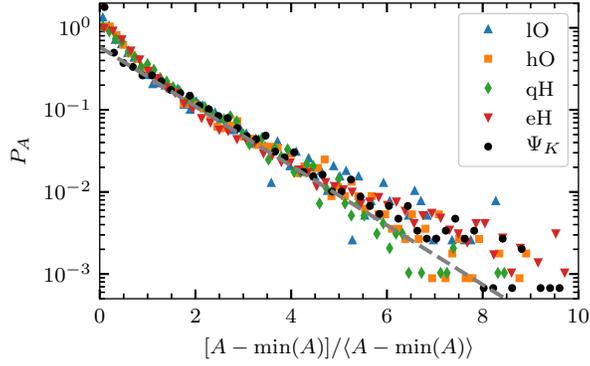}
    \caption{\label{fig:pdf-A} PDF of pulse amplitudes estimated from the deconvolution algorithm for various plasma parameters and confinement states. The grey dashed line gives an exponentially decaying function.}
    % The black circles give the same PDF for a realization of the FPP and the grey dashed line is an exponential function.}
\end{figure}

The pulse amplitude distribution is presented in \Figref{fig:pdf-A} for all plasma parameters and confinement modes, as well as for the synthetically generated signal. These PDFs correspond closely to an exponential distribution over more than two decades in probability. Note that the excess probability for small amplitudes is also present in the synthetically generated signal. From these distributions, we find that $\mA$ for the reconstructed time series is $0.14$ (lO), $0.07$ (hO), $0.24$ (qH) and $0.19$ (eH). For the data time series, $\mA$ can be estimated as $\mA = (\mSnn-\mu)/\gamma$. Using the values from \Tabref{tab:est-par}, we find that $\mA$ for the data time series is $0.13$ (lO), $0.06$ (hO), $0.23$ (qH) and $0.18$ (eH). The deconvolution is highly consistent with estimation using the moments of the data time series.

\begin{figure}
  \centering
    \includegraphics[width=8.5cm]{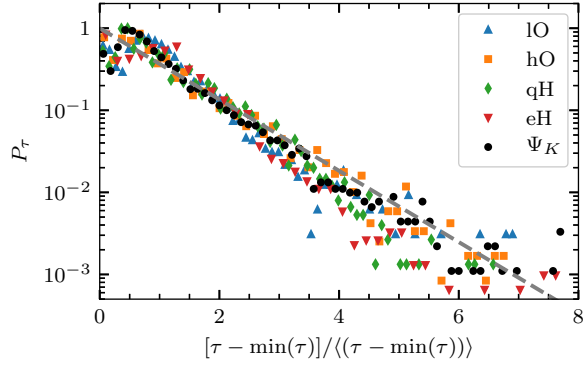}
    \caption{\label{fig:pdf-tw} PDF of waiting times between pulses estimated from the deconvolution algorithm for various plasma parameters and confinement states. The grey dashed line gives an exponential distribution.}
    % The black circles give the same PDF for a realization of the FPP and the grey dashed line is an exponential function.}
\end{figure}

In \Figref{fig:pdf-tw}, the waiting time distribution is presented for all plasma parameters and confinement modes, as well as for the synthetically generated signal. The gray dashed line gives an exponential distribution. All distributions follow an exponential distribution for long waiting times, and the deviation from the exponential distribution for short waiting times is shared by the synthetically generated signal. The average waiting time for these distributions (in $\mu \text{s}$) is $33$ (lO), $12$ (hO), $14$ (qH) and $6.4$ (eH). For the data time series, $\tw$ can be estimated as $\tw = \td/\gamma$. Using $\gamma$ from \Tabref{tab:est-par} and $\td=20\,\mu \text{s}$, we find that $\tw$ for the data time series is $33$ (lO), $12$ (hO), $13$ (qH) and $6.1$ (eH). Again these results are highly consistent.

Using conditional averaging, exponential amplitude and waiting time distributions were found for the same data set; compare \Figsref{fig:pdf-A} and \ref{fig:pdf-tw} in this contribution and Figs. 9 and 10 in \cite{garcia-pop-2018}. Deconvolution provides two advantages over the conditional average: first, the number of found events is one to two orders of magnitude higher, giving clearer distributions over more decades in probability. Secondly, the moments of the deconvolved amplitudes and waiting times can be used directly for comparison with the moments of the original time series. This is not in general possible for the conditionally averaged events.

%In \Figref{fig:pdf-tw}, the waiting times are shown to belong to the same distribution family for all data sets. The dashed grey line shows an exponential function. While the waiting time distributions at large values clearly follow an exponential, there is a lower probability for short waiting times in all cases, even for the synthetically generated FPP with noise. However, neither this lower probability for short waiting times, nor the higher probability for small amplitudes is present in synthetically generated signals without added noise. We therefore conclude that these deviations from the exponential functions are not due to non-exponential waiting times or amplitudes in the data sets.

\begin{figure}
  \centering
    \includegraphics[width=8.5cm]{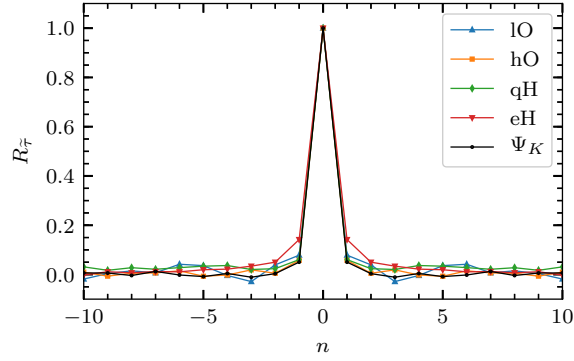}
    \caption{\label{fig:acorr-tw} Auto-correlation function of waiting times between pulses estimated from the deconvolution algorithm for various plasma parameters and confinement states.}
    % The black circles give the result for a realization of the FPP.}
\end{figure}

\begin{figure}
  \centering
    \includegraphics[width=8.5cm]{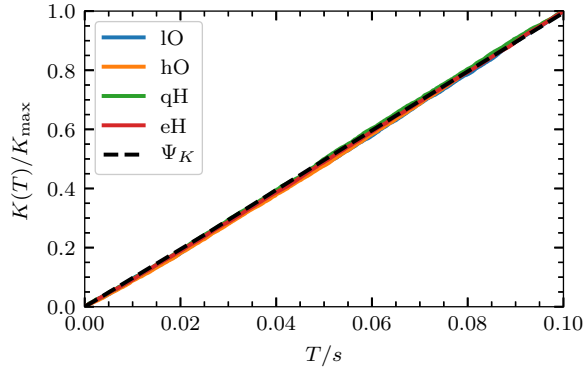}
    \caption{\label{fig:count-tw} Number of pulse arrivals as function of time estimated from the deconvolution algorithm for various plasma parameters and confinement states.}
    % The black dotted line gives the result for a realization of the FPP.}
\end{figure}

The auto-correlation function of the consecutive waiting times is presented in \Figref{fig:acorr-tw}, where $R_{\widetilde{\tau}}[n] = R_{\widetilde{\tau}}[k,k+n] = \langle \widetilde{\tau}_{k}\, {\widetilde{\tau}}_{k+n}  \rangle$. As this is a delta function, consecutive waiting times are uncorrelated and therefore independently distributed. In \Figref{fig:count-tw}, the number of arrivals $K$ as a function of duration $T$ is presented for all data sets. This follows a linear function, showing that the mean value of $K$ can be written as $\tmean{K(T)} = T/\tw$, consistent with a homogeneous Poisson process.

The assumptions of the FPP model are that the number of arrivals follow a homogeneous Poisson process with constant average waiting time $\tw$. Using that the waiting times are independent and that for $\tau>\tmean{\tau}$, they are exponentially distributed, it follows that the process $f_K(t)$ has independent increments and that the number of arrivals $K(T)$ is Poisson distributed. The linearity of $K(T)$ shows that $\tw$ is constant in time. Thus, the process $K(T)$ is a Poisson process with a constant rate of arrivals.

\begin{figure}
  \centering
    \includegraphics[width=8.5cm]{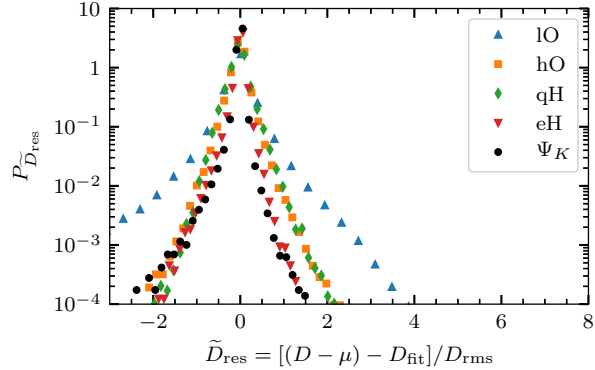}
    \caption{\label{fig:pdf-res} Probability density functions of the residual of the deconvolution algorithm for various plasma parameters and confinement states.}
\end{figure}

Denoting the reconstructed signals as $D_\text{fit}$, the \emph{residual} $D_\text{res}=(D-\mu)-D_\text{fit}$ contains both the error in the reconstruction, as well as the parts of the time series not describable by the FPP. In \Figref{fig:pdf-res}, the PDFs of the residuals is presented, normalized by the rms-value of the original signal such that \Figref{fig:pdf-res} can be directly compared to \Figref{fig:pdf}. These distributions are all sharply peaked and mostly symmetric around the zero-value. The low-density ohmic case is broader than the other distributions, reflecting more pronounced over- and under-estimation of large fluctuations. This difference may be tied to the higher intermittency of the low-density ohmic case compared to the other cases.
For highly intermittent signals, individual deviations from the average exponential shape of the bursts are more pronounced, and so estimation of single pulses is more variable.
Note that none of the distributions are normally distributed, and all seem to follow the same type of distribution as the residual from the synthetic signal. In \cite{theodorsen-deconv}, it will be argued that due to the exponential amplitude distributions and the Gamma distribution of the FPP, normally distributed residuals are not to be expected.

%The agreement between the mean value of $(D-\mu)$ and $D_\text{fit}$ is very good ($\tmean{D_\text{res}}<0.05 D_\rms$), as evidenced by the peak of the mostly symmetric distributions being close to zero.

% - Good agreement with mean value
% - Low standard deviation for all (0.4, 0.2, 0.2 and 0.1 times the standard deviation of the original signals)
% - Some skewness, particularly towards negative values (indicating overestimation of values) (-1.6,0.2,0.19,-1.8).
% - The broader shape of the residual of lO may be tied to its higher intermittency, and reflects the over- and under-estimation in \Figref{fig:detail-data-fit}.
% - Shape is not Gaussian, not even for the synthetic signal. This is not necessarily unexpected \cite{theodorsen-deconv}.

\begin{figure}
  \centering
    \includegraphics[width=8.5cm]{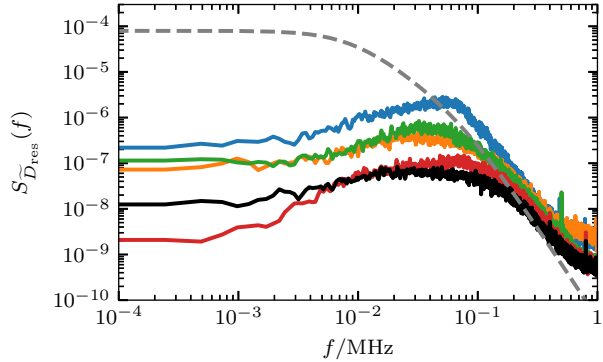}
    \caption{\label{fig:psd-res} Power spectral densities of the residual of the deconvolution algorithm for various plasma parameters and confinement states. The color coding is the same as in \Figref{fig:psd}, the black full line gives the residual of the synthetic signal $\Snn_K$ and the gray dashed line gives the spectrum of the FPP with pulse parameters $\td = 20\,\mu\text{s}$ and $\lambda=1/10$.}
\end{figure}

 In \Figref{fig:psd-res}, the PSD of the reconstructed time series is presented. For low frequencies, below about $10^{-2}\, \text{MHz}$, the power density is very moderate, below $1\%$ of the power contained in comparable frequencies for the data time series, \Figref{fig:psd}. On the other hand, the power content at high frequencies, above about $10^{-1}\,\text{MHz}$, is comparable to that of the data time series as a whole. This may be due to high-frequency noise filtered out by the deconvolution algoritm. The residual of the low-density ohmic case contains the most power, consistent with the broader PDF in \Figref{fig:pdf-res}.

\section{Conclusions}
The FPP with exponentially shaped pulses and exponentially distributed pulse amplitudes has previously successfully predicted all statistical properties of SOL fluctuations as recoded by single-point measurement. This comprises the amplitude PDF \cite{graves-2005,garcia-jnm-2013,garcia-pop-2013,garcia-nf-2015,kube-2016,theodorsen-ppcf-2016,garcia-nme-2017,theodorsen-nf-2017,garcia-pop-2018}, the auto-correlation function and the frequency power spectral density \cite{theodorsen-ppcf-2016,garcia-nme-2017,theodorsen-nf-2017,garcia-pop-2018} and level crossing rates and excess time statistics \cite{garcia-nme-2017,theodorsen-nf-2017}. In this contribution, a deconvolution algorithm is used in order to directly and unambiguously recover pulse amplitudes and arrival times, verifying the underlying assumptions of the stochastic model.

This algorithm is applied to GPI data time series that recorded emission fluctuations in the SOL of the Alcator C-Mod device for various plasma parameters and confinement modes. The statistical properties of far-SOL fluctuation arrival times and amplitudes have been shown to be the same in all cases. Both the pulse amplitudes and waiting times are exponentially distributed. Moreover, the waiting times are uncorrelated and the number of pulse arrivals increases linearly with the time series duration. This demonstrates that the statistics of far SOL fluctuations are the same for ohmic and H-mode plasmas in the Alcator C-Mod device, and in particular that the pulses arrive according to a homogeneous Poisson process and have exponentially distributed amplitudes, justifying all the assumptions underlying the stochastic model. This provides strong evidence in support of universal applicability of the stochastic model, providing a valuable tool for describing intermittent fluctuations and associated plasma--wall interactions in the boundary region of magnetically confined plasmas. The properties of the deconvolution algorithm will be elucidated in \cite{theodorsen-deconv}.

\begin{acknowledgements}
This work was supported with financial subvention from the Research Council of Norway under grant 240510/F20 and the U.S. Department of Energy, Office of Science, Office of Fusion Energy Sciences, using User Facility Alcator C-Mod, under Award Number DE-FC02-99ER54512- CMOD. A.~T., O.~E.~G. and R.~K. acknowledge the generous hospitality of the MIT Plasma Science and Fusion Center where parts of this work was conducted.
\end{acknowledgements}

% When .makepdf.sh is run, the source file is used to generate a .bbl file. This file is then used afterwards, to make the actual pdf-output.
% When submitting the article, first generate the .bbl with makepdf.sh, then swich commenting and use pdflatex twice to mirror the process used by the journal.
\IfFileExists{theodorsen-v3.bbl}{\input{theodorsen-v3.bbl}}{\bibliography{sources}\bibliographystyle{aipnum4-1}}

\end{document}